# Intersubband Transitions in Lead Halide Perovskite-Based Quantum Wells for Mid-Infrared Detectors

*Xinxin Li[1,2], Wanyi Nie[3], Xuedan Ma[1,2]*

[1]Center for Nanoscale Materials, Argonne National Laboratory, Lemont, Illinois 60439, United States

[2]Consortium for Advanced Science and Engineering, University of Chicago, Chicago, Illinois 60637, United States

[3]Center for Integrated Nanotechnologies, Los Alamos National Laboratory, Los Alamos, New Mexico 87545, United States

ABSTRACT

Due to their excellent optical and electrical properties as well as versatile growth and fabrication processes, lead halide perovskites have been widely considered as promising candidates for green energy and opto-electronic related applications. Here, we investigate their potential applications at infrared wavelengths by modeling the intersubband transitions in lead halide perovskite-based quantum well systems. Both single-well and double-well structures are studied and their energy levels as well as the corresponding wavefunctions and intersubband transition energies are calculated by solving the one-dimensional Schrödinger equations. By adjusting the quantum well and barrier thicknesses, we are able to tune the intersubband transition energies to cover a broad range of infrared wavelengths. We also find that the lead-halide perovskite-based quantum wells possess high absorption coefficients, which are beneficial for their potential applications in infrared photodetectors. The widely tunable transition energies and high absorption coefficients of the perovskite-based quantum well systems, combined with their unique material and electrical properties, may enable an alternative material system for the development of infrared photodetectors.

**Introduction**

Spectroscopy in the thermal infrared (IR) region spanning the wavelength range of 3 – 15 μm is critical for many applications ranging from astronomy to security to environmental analysis. Detection of IR radiation could be implemented based on thermal effects like those in bolometers[1] or interband transitions in narrow gap ternary alloys such as mercury cadmium telluride.[2] An alternative approach to these technologies replies on the intersubband (ISB) transitions in epitaxially grown nanostructures such as quantum wells (QWs) and quantum dots (QDs).[3-6] The versatility in the compositions and electronic structures of these nanostructures allows precise tuning in their ISB transition energies and rates through the quantum confinement effect. So far, IR light sources and detectors such as quantum cascade lasers and detectors based on ISB transitions in quantum wells and quantum dots have been successfully demonstrated and utilized in many important applications.

Despite these developments in thermal IR devices based on epitaxially grown nanostructures, they remain technologically demanding and mostly limited to defense and scientific applications. Low-cost alternatives combining inexpensive material synthesis and ease of processing is highly desirable[7] and could potentially offer practical solutions to sophisticated device designs and their wider applications. In this regard, perovskite materials may serve as a cost-effective alternative. Specifically, their tunable compositions and versatile growth and fabrication procedures may allow the integration of multi-spectral components in a single device. Although the current generation of perovskite materials have exhibited moderate mobilities at room temperature when compared to some of the best performing inorganic semiconductors such as GaAs, their long diffusion lengths and charge recombination lifetimes ensure excellent charge extraction performance.[8] Moreover, perovskite materials possess benign defect properties where deep defect formation is energetically unfavorable,[9] indicating that high quality devices based on perovskites can be fabricated with high tolerance to imperfections.[10] This is in stark contrast to common QW materials such as GaAs, which typically require high purity and crystallinity to obtain high device performance. Given the encouraging progresses in perovskite material design and growth, it is expected that higher carrier mobilities would be achieved in perovskites incorporating light metals.[11] These properties together position perovskites as a potentially compelling material system for developing IR devices.

Aside from these appealing material properties of perovskites, recent developments in their thin film growth and controlled doping may further facilitate their applications in IR devices. Ultrathin perovskite films that are a few nanometer thick have been grown using various approaches[12-14] and interband emission from such perovskite-based QWs has been demonstrated.[12] Although efficient and reliable tuning of carrier densities in perovskites has been proved to be challenging, encouraging progresses have been achieved through chemical composition and crystal structure engineering.[15-16] These developments provide necessary foundations for the implementation of IR devices based on perovskites.

Here, we explore theoretically QW structures based on perovskite materials that would allow ISB transitions suitable for IR applications. Using cesium lead halide perovskites (CsPbX$_3$) as an illustrative case, we start by investigating the energy levels and ISB transitions in perovskite-based single quantum wells (SQWs) and then continue on to symmetric double quantum wells (DQWs). ISB transitions corresponding to IR wavelengths were achieved by tuning the well and barrier thicknesses. We demonstrate that ISB absorption coefficient as high as $10^4$ cm$^{-1}$ can be achieved in the 5 – 11 μm wavelength range in CsPbX$_3$-based SQWs and DQWs, making them promising materials for IR photodetectors.

**Simulation Methods**

We use cesium lead halides (CsPbX$_3$) as the prototype perovskite materials for the design of QWs for IR applications. We expect similar approaches and conclusions can be extended to other types of inorganic perovskite materials. One-dimensional (1D) Schrödinger equations were solved numerically using the software package *nextnano$^3$*.[17] Due to the relatively large bandgaps of the perovskite materials, the coupling between their conduction and valence bands is ignored.[18] The single-band Schrödinger equation can be expressed as:[19]

$$\boldsymbol{H}\Psi_n(\boldsymbol{r}) = E_n \Psi_n(\boldsymbol{r}) \qquad (1)$$

$$-\frac{\hbar^2}{2}\nabla \cdot \boldsymbol{M}(\boldsymbol{r})\nabla + V(\mathbf{r})\Psi_n(r) = E_n \Psi_n(r) \qquad (2)$$

with V(r) being the potential energy, and the effective mass tensor $\boldsymbol{M}(x)$ as:

$$\begin{pmatrix} 1/m_{xx} & 1/m_{xy} & 1/m_{xz} \\ 1/m_{yx} & 1/m_{yy} & 1/m_{yz} \\ 1/m_{zx} & 1/m_{yz} & 1/m_{zz} \end{pmatrix} \qquad (3)$$

The 1D single-band effective mass Schrödinger equation is written as:

$$\left(-\frac{\hbar^2}{2m_{xx}} + V(\mathbf{x})\right)\Psi_n(x) = E_n\Psi_n(x) \tag{5}$$

For semiconductors with periodic potential: $V(\mathbf{x}) = V(\mathbf{x} + \mathbf{L})$ in which $\mathbf{L}$ is the length of periods, the eigenfunctions can be expressed as:

$$\Psi_{n\mathbf{k}}(x) = u_{n\mathbf{k}}(x)e^{i\mathbf{k}\cdot x} \tag{6}$$

With $u_{n\mathbf{k}}(x) = u_{n\mathbf{k}}(x + \mathbf{L})$. In the multi-band $\mathbf{k} \cdot \mathbf{p}$ method, the relativistic effect of spin is taken into consideration by adding an additional term $\mathbf{H}_{so}$ which has the form of:

$$\mathbf{H}_{so} = \frac{\hbar^2}{4m_0^2 c^2}(\nabla V \times \mathbf{p}) \cdot \sigma \tag{7}$$

in which $m_0$ is the electron mass, $c$ is the speed of light in vacuum, $\mathbf{p}$ is the momentum operator, and $\sigma$ is the vector of the Pauli matrices. Taken together, the Schrödinger equation can be written as:

$$(\mathbf{H_0} + \mathbf{H}_{so})\Psi_{n\mathbf{k}}(x) = E_n(\mathbf{k})\Psi_{n\mathbf{k}}(x) \tag{8}$$

After inserting the Bloch function $\Psi_{n\mathbf{k}}(x)$ and cancelling $e^{i\mathbf{K}\cdot x}$ term, one obtains:

$$\left(\frac{p^2}{2m_0} + V(x) + \frac{\hbar^2 k^2}{2m_0} + \frac{\hbar}{m_0}\mathbf{k}\cdot\mathbf{p} + \frac{\hbar^2}{4m_0^2 c^2}(\nabla V \times \mathbf{p})\cdot\sigma\right)u_{n\mathbf{k}}(x) = E_n(\mathbf{k})u_{n\mathbf{k}}(x) \tag{9}$$

Electronic structures were represented within both the single-band and eight-band effective mass function approximations with periodic potentials. The wavefunctions, energy levels and band edges along the QW depth were calculated at room temperature in this study. We use eight-band $\mathbf{k} \cdot \mathbf{p}$ Schrödinger equation to derive the wavefunctions, energy levels and interband transition energies of the QW structures unless otherwise stated. The probability of the ISB transitions between an initial state $i$ and a final state $f$ can be obtained by evaluating the ISB transition dipole moments $z_{fi}$:[20-21]

$$|z_{fi}| \propto \left| \int \psi_f^*(z) \times z \times \psi_i(z) dz \right| \qquad (10)$$

with $\psi_f^*$ and $\psi_i$ being the wavefunctions of the initial and final states, and z the direction perpendicular to the QW layers.

For absorption coefficients in doped structures, Poisson equation was also included to reflect charge self-consistence:[19]

$$\nabla \cdot (\varepsilon_0 \varepsilon_r(\mathbf{x}) \nabla \phi(\mathbf{x})) = -\rho(\mathbf{x}) \qquad (11)$$

where $\varepsilon_0$ is the vacuum permittivity, $\varepsilon_r(\mathbf{x})$ is the static dielectric constant tensor at position $\mathbf{x}$ and $\phi$ is the electrostatic potential. Based on the ISB transition dipole matrix element $z_{fi}$, we further calculate the absorption coefficient using:[20, 22]

$$\alpha(\omega) = \frac{e^2 \omega (N_i - N_f)}{\hbar \varepsilon_0 n c} z_{ij}^2 \frac{\Gamma/2}{(E_j - E_i - \hbar\omega)^2 + \Gamma^2/4} \qquad (12)$$

where $c$ is the speed of light in vacuum, $n$ is the refractive index, and $\Gamma$ is the energy linewidth in terms of full-width-half-maximum (FWHM) which is derived as $\Gamma = \hbar/\tau$ with $\tau$ being the relaxation time. $N_i$ is the volume electron density which can be derived from the 2D density of states and Fermi-Dirac function:[20-21]

$$N_i = \frac{m^* k_B T}{\pi \hbar^2} \ln\left(\frac{1}{1-f_i}\right) = \frac{m^* k_B T}{\pi \hbar^2} \ln\left(1 + \exp\left(\frac{E_f - E_i}{k_B T}\right)\right) \qquad (13)$$

where $k_B$ is the Boltzmann constant, $T$ is the temperature and $f_i$ is the occupancy probability of the $i$th band based on the Fermi-Dirac distribution.

Even though reports have demonstrated the coexistence of orthorhombic and cubic phase in nanocrystalline CsPbBr3 at room temperature [23, 24], during our simulation, we consider a simplified situation and assume the crystal structure to be cubic, as illustrated in Figure 1. Cubic ABX$_3$ perovskites have direct bandgaps at the R point (½, ½, ½) of the Brillouin zone.[25] Figure 2(a) shows the representative band structure of cubic CsPbX$_3$ perovskites. The top of their valence band (blue curve in Figure 2(a)) is $s$-like which is doubly degenerated considering the spin degeneracy. Due to the strong spin-orbit coupling (SOC) in CsPbX$_3$ perovskites, the well-known Pb 6$p$ orbitals converts to a singlet state at the bottom of conduction band (pink curve in Figure 2(a)) and a three-fold degenerated triplet state with the degeneracy multiplying by 2 considering spin.[25] The top two conduction bands in Figure 2(a) represent the light electron (LE) and heavy electron (HE) states, respectively, while the bottom of the conduction band is the split-off (SO) state with a spin-split off energy $\Delta_{SO}$ = 1.5 eV for CsPbBr$_3$.[26] This is inverted from the band structures observed in common III-V semiconductors (Figure 2(b)) which consists of a *s*-like conduction band and a valence band comprised of a heavy hole (HH), a light hole (LH) and a SO state.

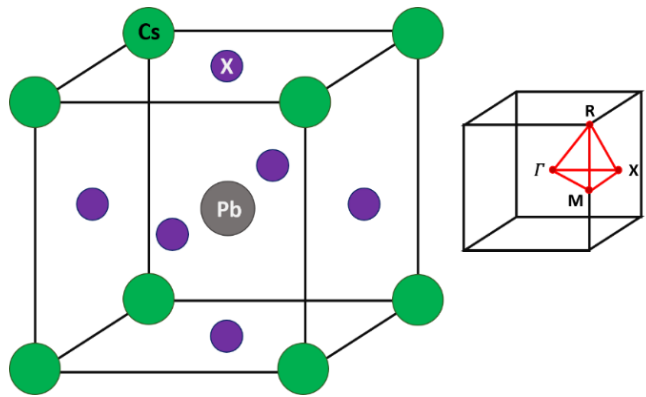

**Figure 1**. The crystalline structure for cubic CsPbX$_3$ perovskite. The inset is an illustration of the Brillouin zone for the cubic structure.

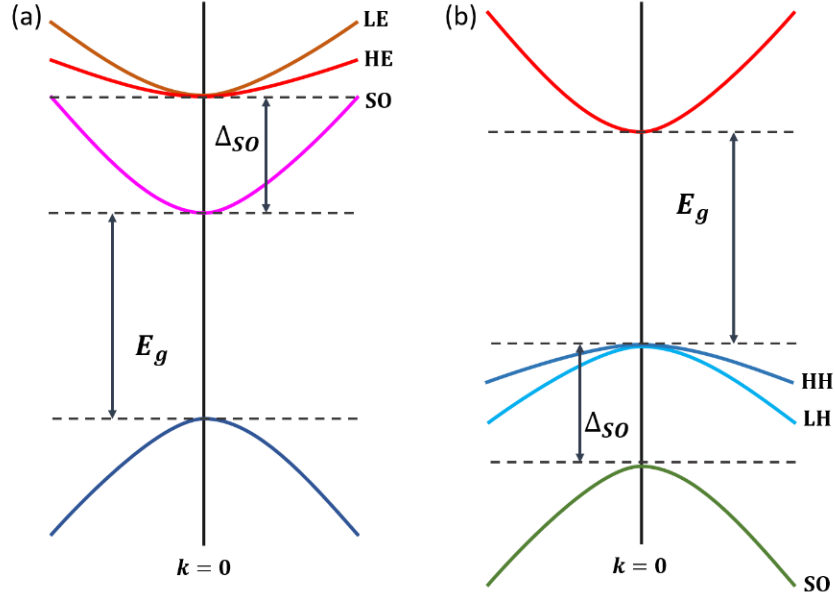

**Figure 2.** Band edge energy levels in (a) CsPbX$_3$ perovskites and (b) common III-V semiconductors.

The eigenvectors of the effective mass tensor are oriented along the Pb-Pb directions which are all equivalent for cubic structure.[27] We assume both the hole and electron masses to be isotropic in our simulation. Table 1 lists the major parameters, including the lattice constants, band gaps ($E_g$), effective masses, Luttinger parameters, dielectric constants, and phonon energies of the CsPbX$_3$ materials used in this study. Solution-grown CsPbX$_3$ perovskites normally show slightly p-type behavior[28] with hole concentration less than $10^{14}$ cm$^{-3}$.[29] Given this low doping concentration, we assume the as-grown CsPbX$_3$ perovskites to be undoped in the our calculations unless otherwise mentioned.

**Table 1.** Parameters used for the calculations of the energy band diagrams and wavefunctions.

| Parameters | CsPbBr$_3$ | CsPbCl$_3$ |
|---|---|---|
| Lattice constant (Å) [30-31] | 5.77 | 5.62 |
| E$_g$ (eV) [26, 30] | 2.34 | 3.04 |
| $m_h(m_0)$ [26, 32] | 0.26 | 0.17 |
| $m_{he}(m_0)$ [26] | 1.605 | 1.619 |
| $m_{le}(m_0)$ [26] | 0.303 | 0.357 |
| $m_{SO}(m_0)$ [26] | 0.509 | 0.584 |
| $\Delta_{SO}$ (eV) [25] | 1.5 | 1.54 |
| $\gamma_1^L$ [26] | 7.6 | 5.8 |
| $\gamma_2^L$ [26] | 3.0 | 2.2 |
| $\gamma_3^L$ [26] | 0.7 | 0.5 |
| Static dielectric constant $\varepsilon_s$ [33] | 15.6 | 15.6 |
| High frequency dielectric constant $\varepsilon_\infty$ [33-34] | 4.96 | 4.5 |
| LO phonon Energy (meV) [35-36] | 18 | 42.8 |

## Results and Discussion

### Interband Transitions in CsPbBr$_3$ Single Quantum Wells

We first consider SQW structures based on CsPbX$_3$ perovskites. Lee *et al*[12] have recently reported interband transitions at visible wavelengths from CsPbBr$_3$-based SQWs with 1,3,5-tris(N-phenylbenzimidazol-2-yl)benzene (TPBi) as the barriers. To benchmark our simulation, we start by calculating the interband transitions in a similar SQW structure and comparing the results to the reported values. The valence and conduction band offset energies are set to be 0.53 eV and 0.32 eV, respectively, which are derived from the band alignment between CsPbBr$_3$ and TPBi.[12] The width of the CsPbBr$_3$ SQW is varied from 2 to 20 nm while that of the TPBi barrier is fixed to be 7 nm.

**Figure 3**(a) shows the simulated wavefunctions and energy levels of a representative 5 nm thick CsPbBr$_3$ SQW embedded in TPBi barriers. From top to down, the bands are the valence band $E_V$ and conduction band $E_C$, respectively. Due to the large band offsets between the QW and the barriers, both the electrons and holes are highly confined in the QW region. A reasonable agreement between the bandgaps obtained from our simulations and those derived from the experimental data in Ref. 12 can be observed (Figure 3(b)). For the QWs with thicknesses smaller than the Bohr radius of CsPbBr$_3$ (~ 7 nm),[37] a prominent quantum confinement effect is expected,

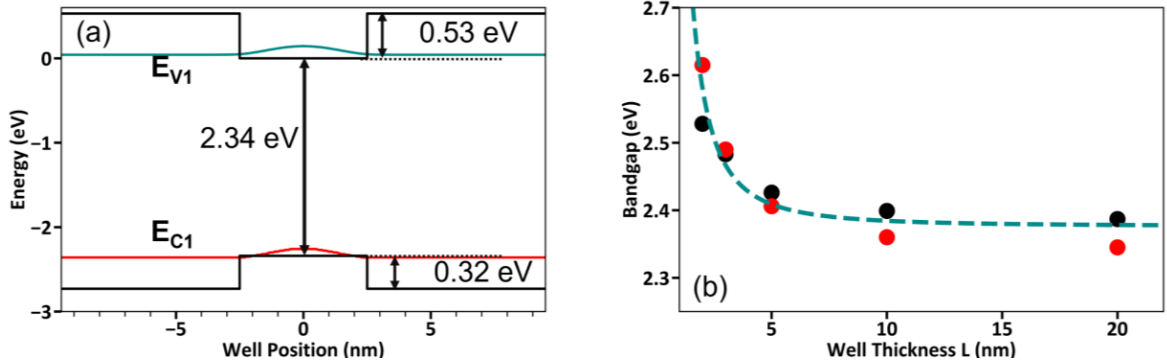

**Figure 3.** (a) Band alignment and wavefunctions of a 5 nm thick single CsPbBr$_3$ quantum well surrounded by 7 nm thick TPBi barriers. $E_{V1}$ and $E_{C1}$ are the lowest energy levels in the valence and conduction bands, respectively. (b) Calculated bandgap energies of CsPbBr$_3$/TPBi single quantum well/barrier structures with varying well thicknesses (red dots) compared to those derived from the experimental data in Ref. 12 (black dots). The cyan curve is a fit of the data using the "particle-in-a-box" model.

as is evident from Figure 3(b). The bandgap increases from around 2.4 eV for a 10 nm thick SQW to ~2.6 eV for a 2 nm thick SQW, a trend that can be well described by the simple "particle-in-a-box" model which assumes that the interband transition energy is inversely proportional to the carrier effective mass (Figure 3(b), cyan curve).[12] These simulation results attest the accuracy of the eight-band $\mathbf{k} \cdot \mathbf{p}$ approach, which we will apply primarily in our following calculations of the intersubband transitions in the perovskite quantum well structures.

**Intersubband Transitions in CsPbBr$_3$ Single Quantum Wells**

To investigate the ISB transitions in perovskite QWs, we use the widely studied CsPbBr$_3$ as the quantum well material for illustrative purposes, although we expect similar approaches can be applied to other perovskite systems. For simplicity, CsPbCl$_3$ is employed as the barrier due to their large band offsets compared to CsPbBr$_3$ and widely available material parameters. Same as in the case of the QW materials, in practice CsPbCl$_3$ can be replaced by other materials as long as they meet the band alignment requirements to serve as the barriers.

In the CsPbBr$_3$/CsPbCl$_3$ QW/barrier heterostructure, due to the non-negligible 2.5% lattice mismatch between CsPbBr$_3$ and CsPbCl$_3$, we estimate the critical thickness for CsPbBr$_3$ to grow on CsPbCl$_3$ with dislocation free transition to be around 9 nm[38] by assuming a Poisson's ratio of around 0.28.[39] Therefore, in our simulations, the thickness of the CsPbBr$_3$ SQW is kept below the critical thickness, while the CsPbCl$_3$ barrier thickness is fixed to 10 nm. The absolute conduction and valence band edge energy levels were reported to be -3.26 eV and -6.24 eV for CsPbCl$_3$,[40] and -3.6 eV and -5.9 eV for CsPbBr$_3$,[41] which give rise to around the same conduction and valence offsets of 0.34 eV.

Figure 4(a) depicts the band structure of a representative 5 nm thick CsPbBr$_3$ SQW sandwiched between two 10 nm CsPbCl$_3$ barrier layers. From top to down, the bands correspond to the valence, SO and HE/LE states, respectively. Figure 4(b) shows the wavefunctions of the three lowest states, namely the ground state $|1\rangle$, the first and the second excited states $|2\rangle$ and $|3\rangle$, that are confined in the valence band of the 5 nm thick SQW. Much like that in III-V QWs, the ISB transitions in the perovskite QWs are only allowed between states with same parity. The ISB transitions that are allowed by the optical selection rules have energies of $E_{32} = E_3 - E_2 = 120.6$ meV and $E_{21} = E_2 - E_1 = 85.4$ meV, respectively. We obtained corresponding ISB transition dipole moments of $\boldsymbol{z_{32}} = 1.35$ nm and $\boldsymbol{z_{21}} = 1.21$ nm, respectively, which are comparable to those commonly obtainable in III-V QWs.[42]

One of the major advantages offered by ISB transitions are their flexibility in tailoring the transition energies through tuning the quantum well thickness. We systematically investigate the influence of the quantum well thickness on the ISB transition energies and the simulation obtained results are shown in Figure 4(c). As shown in Figure 4(c), when the quantum well becomes thinner, the ISB transition energies $E_{32}$ and $E_{21}$ increase accordingly. Changes in the SQW thickness also affect the electronic states that can be confined in the SQWs. When the SQW is thinner than 4.0 nm, the energy level of the second excited state $|3\rangle$ becomes higher than that of the barrier. Consequently, only the ground state $|1\rangle$ and the first excited state $|2\rangle$ can be confined in the SQWs (see Figure S1 for an example) and only $E_{21}$ can be obtained (Figure 4(c)). In contrast, when the SQW is thicker than 5.5 nm, the energy level of the third excited state $E_4$ becomes lower than the barrier height and it becomes confined in the SQW. By adjusting the SQW thickness, we are able to tune the ISB transition energies from 235 meV to 66 meV, covering most of the thermal IR range.

**Figure 4**(d) illustrates the normalized absorption coefficient for intersubband transition $T_{1->2}$ in SQWs with well thicknesses varying from 2.0 to 2.8 nm. To bring the Fermi level above $E_1$, we deliberately p-dope the central region of the SQW for a symmetric doping profile. The doped region is fixed to be 9 nm regardless of the well thickness in order to keep the doping profile consistent. Since the absorption coefficient is highly dependent on the doping concentration,[43] we

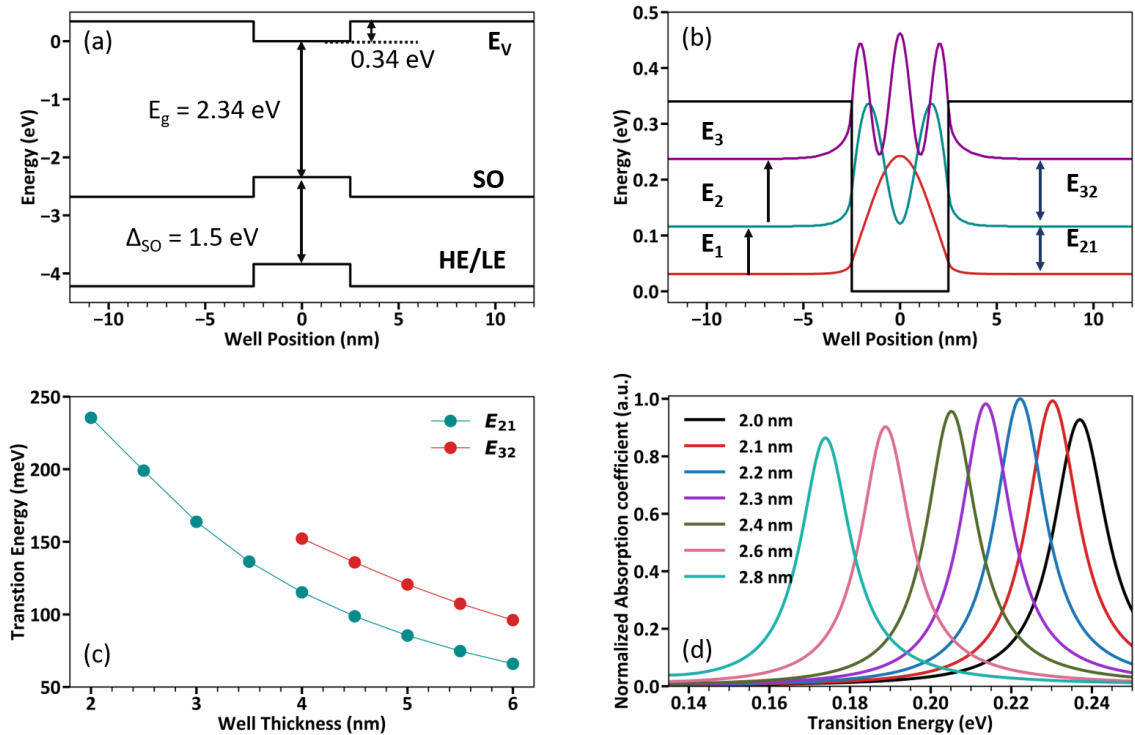

**Figure 4** (a) schematic illustration of the energy bands in a 5 nm thick SQW, (b) the wavefunctions of the three confined states in the valence band, (c) transition energy $E_{32}$ and $E_{21}$ versus well thickness, (d) normalized absorption coefficient for $T_{1->2}$ in SQW structure with varying well thickness.

assume the same doping level of $3\times10^{18}$ cm$^{-3}$. for all the SQWs. We note that the doping level we adapt here for the simulations is already achievable experimentally,[44] although more studies are required for better controllability. In Figure 4(d), it is noticeable that the absorption coefficient first increases with the reducing well thickness and then decreases, which is consistent with previous reports on III-V SQWs.[21] The maximum absorption coefficient (~5100 cm$^{-1}$ with doping concentration of $3\times10^{18}$ cm$^{-3}$ in Figure S2) occurs at the well thickness of around 2.2 nm (Figure 4(d)), which corresponds to an $E_{21}$ value of around 220 meV. This implies that an optimal absorption efficiency of ISB transition photodetectors based on CsPbBr$_3$ SQWs is likely to occur at a working wavelength of around 5.6 μm. We also investigate the influence of the doping level and observe the maximum achievable absorption coefficient of the SQWs increases consistently with the doping level (Supporting information S2). Moreover, aside from the tunable operation wavelength spanning 5 - 7 μm, the simulation obtained nominal absorption coefficients of the studied SQWs are relatively high (~13000 cm$^{-1}$ for $1\times10^{19}$ cm$^{-3}$), comparable to those of heavily doped GaAs SQW (~ 6000 cm$^{-1}$ for $5\times10^{19}$ cm$^{-3}$) and nonpolar GaN SQW (~10000 cm$^{-1}$ for ~$1\times10^{20}$ cm$^{-3}$).[45] The large ISB absorption coefficients, combined with the long charge carrier diffusion lengths and defect tolerant properties of the perovskite materials, may potentially offer an alternative approach for high performance ISB transition photodetectors.

**Intersubband Transitions in Symmetric CsPbBr$_3$ Double Quantum Well (DQW)**

Compared to SQWs, multiple-well structures provide wider energy tunability due to the possibility to separate the charge carriers in the different QWs by manipulating the well and barrier thicknesses. Figure 5(a) shows a representative symmetric DQW structure comprised of two identical CsPbBr$_3$ quantum wells both with a well thickness of 2.5 nm and a central CsPbCl$_3$ barrier with a thickness of 1.5 nm. The thicknesses of the side barriers are fixed to be 7 nm. Figure 5(b) illustrates the four energy levels confined in the symmetric DQW. Our numerical calculations yield ISB transition energies of $E_{41} = 231.9$ meV, $E_{32} = 178.2$ meV and $E_{21} = 17.2$ meV, with the latter being close to the longitudinal optical phonon energies in CsPbBr$_3$ (around 18 meV). [46] To investigate the influence of the central barrier thickness, we keep the thickness of both QWs to 2.5 nm and adjust the barrier thickness from 1.0 nm to 2.0 nm. As can be seen in Figure 5(c), an increase in the central barrier thickness reduces E$_{21}$ but increases E$_{32}$. On the other hand, if we keep the central barrier thickness at 1.0 nm and tune the thickness of the QWs, both E$_{32}$ and E$_{21}$ decrease with the increasing central well thickness (Figure 5(d)). These simulation results reveal the flexibility in tuning the transition energies in the DQWs. The functionality of the DQWs can be further extended if asymmetric quantum wells are considered.

We further consider the potential of using the ISB transitions in perovskite CQWs for photodetection applications by simulating the achievable absorption coefficients in these heterostructures. Since the energy levels |1⟩ and |2⟩ are close in the current design, the transitions that are of interest for mid-IR application is from energy level |2⟩ to |3⟩ denoted as T$_{2->3}$. In this case, we again *p*-dope 9 nm of the central region to bring the Fermi level above E$_2$. The doping level is kept at $3\times10^{18}$ cm$^{-3}$. We first investigate the influence of the central barrier thickness on the absorption coefficient by keeping the quantum well thickness fixed at 2.1 nm while tuning the central barrier thickness from 1.2 to 2.4 nm. As can be seen in Figure 5(e), the absorption coefficient for the T$_{2->3}$ transition first increases with the central barrier thickness and then decrease with a maximum occurring at the central barrier thickness of 1.5 nm. We also investigate the influence of the quantum well thickness on the absorption coefficient. By keeping the central barrier thickness to 1.5 nm, the absorption coefficient maximum of around 1400 cm$^{-1}$ is found for a quantum well thickness of 2.2 nm (Figure 5(f)). The influence of the doping concentration on

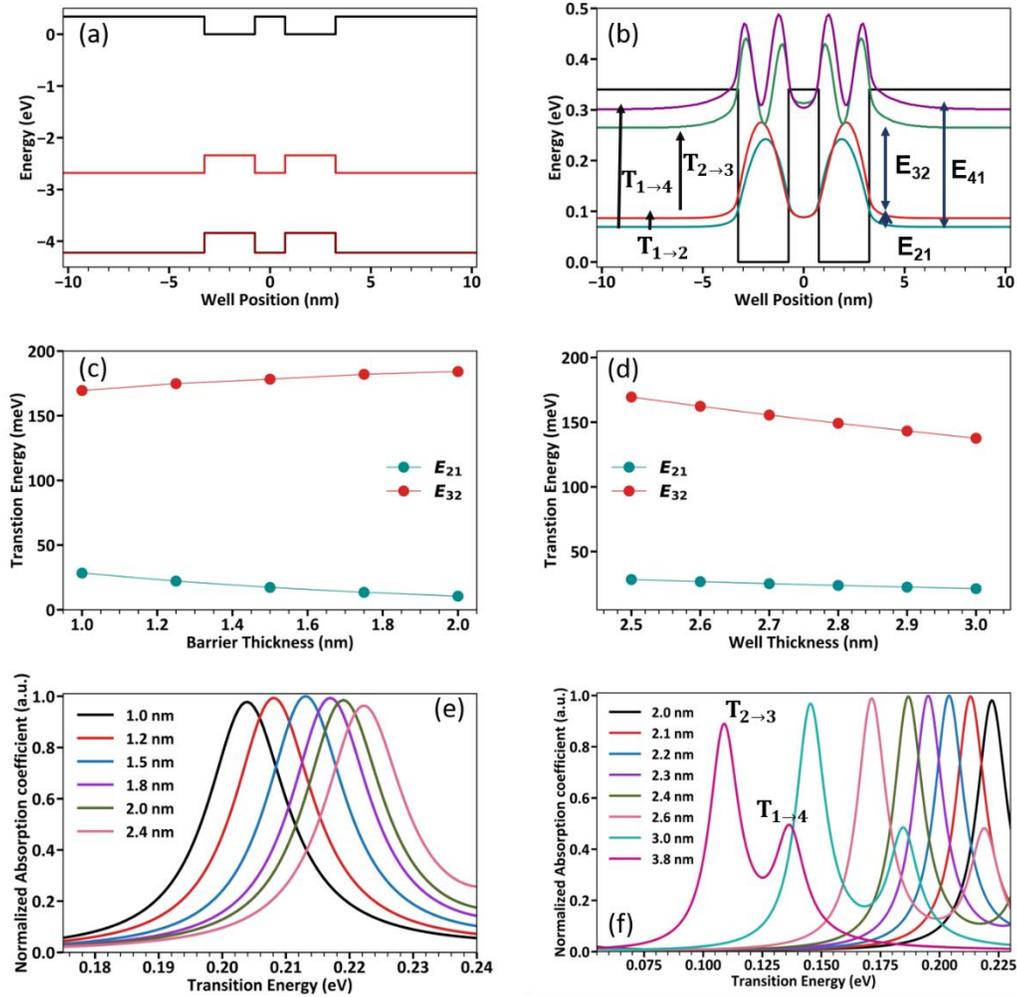

**Figure 5** (a) A illustration of the energy band structure for $CsPbBr_3/CsPbCl_3$ symmetric DQWs. (b) The wavefunctions of the confined states in the valence band of 2.5 nm/ 1.5 nm/2.5 nm DQW obtained from the $\mathbf{k} \cdot \mathbf{p}$ method. (c) The central barrier thickness-dependent transition energies $E_{21}$ and $E_{32}$. (d) The quantum well thickness-dependnent transition energies $E_{21}$ and $E_{32}$. (e) The absorption coefficient of $CsPbBr_3/CsPbCl_3$ symmetric DQWs with central barrier thickness varying from 1.2 to 2.4 nm. The doping level is kept at $3\times10^{18}$ $cm^{-3}$. (f) The absorption coefficient of $CsPbBr_3/CsPbCl_3$ symmetric DQWs with the quantum well thickness varying from 2.0 to 3.8 nm. The low energy peaks correspond to $T_{2\rightarrow3}$ transitions and the higher energy peaks correspond to $T_{1\rightarrow4}$ transitions.

the absorption coefficient is similar to that in the SQWs: an increase in the doping level leads to a larger maximum achievable absorption coefficient. At the doping level of 1×10$^{19}$ cm$^{-3}$, the absorption coefficient can be as high as 12000 cm$^{-1}$ (see supporting information Figure S2(b)) with an optimal working wavelength of ~ 5.6 μm ($E_{23}$ = 223.2 meV). We note that although this absorption coefficient for the T$_{2->3}$ transition in the DQWs is comparable to the T$_{1->2}$ transition in the SQW systems discussed in the previous section with similar high doping profiles, the DQW geometries studied in Figure 5(e) and 5(f) provide high absorption coefficients covering a much wider working wavelength range from around 5 to 11 μm.

Moreover, DQWs could allow decoupling of the ground and excited states in the different QWs, thus afford more tunability both in terms of the covered ISB transition energies and rates. This is of particular importance when designing ISB transitions in QWs for IR light source applications. In these cases, a three-level system is typically used with level |1⟩ being the ground state and |2⟩ and |3⟩ the first and second excited states. When electrons are optically excited or electrically injected into the second excited level, population inversion and simulated emission occurs between the first |2⟩ and second |3⟩ excited states provided that the lifetime of the first excited level |2⟩ is shorter than that of the second excited level |3⟩, which can be achieved by tuning the energy spacing between the subbands to promote or inhibit relevant electron-phonon interactions.[42] Specifically, the optical gain in such a three-level system is determined by:[47] $F \propto (1 - \tau_{21}/\tau_{32}) \cdot \tau_3 \cdot z_{31}^2 z_{32}^2$, where $\tau_{21}$ and $\tau_{32}$ are the ISB transition lifetimes, $\tau_3 = (\tau_{31}^{-1} + \tau_{32}^{-1})^{-1}$ is the lifetime at the second excited state |3⟩, and $z_{31}$ and $z_{32}$ are the ISB transition dipole matrix elements. The scattering lifetime $\tau_{if}$ from an initial state $i$ to a final state $f$ is highly dependent on the electron-phonon interaction overlap integral, where the integration covers the allowed values of optical phonon wave vector. In order to have the first excited state relax much faster than the second excited state to achieve population inversion, i.e. $\tau_{21} \gg \tau_{32}$, the first excited state could be chosen to be spaced by the optical phonon energy from the ground state to promote electron-phonon interaction. In the example case shown in Figure 5(a) and 5(b), the $E_{21}$ transition energy is tuned to be close to the longitudinal optical phonon energies in $CsPbBr_3$. Therefore, this DQW profile may allow the efficient depopulation of the first excited level $|2\rangle$ and facilitate population inversion between the second and first excited states, thus offering optical gains unattainable using SQWs (see Supporting Information S3 for details). We note that while this design principle has been implemented successfully in III-V QWs, the abundance of different phonon modes in perovskites caused by either the motions of the inorganic lead halide ions or the lattice vibrations of the organic cations signify the importance of suppressing electron-phonon interactions in order to obtain high gains in perovskite-based IR photon sources. Recent experimental demonstrations show that the application of inorganic perovskites incorporating light metals can effectively eliminate phonon modes associated with the organic cations and increase the phonon frequencies.[8] The versatile compositions of the perovskite materials may offer promising controls over the gain properties of perovskite-based IR photon sources.

**Conclusions**

We constructed $CsPbBr_3$/$CsPbCl_3$ SQWs and symmetric DQW structures with varying well and barrier thicknesses and calculate their wavefunctions, intersubband transition energies and absorption coefficients. Our simulations show that highly *p*-doped SQWs with a 2.2 nm well thickness provides maximum achievable absorption coefficients (~13000 cm$^{-1}$ at the doping concentration of $1\times10^{19}$ cm$^{-3}$) at a working wavelength of ~ 5.6 μm. The DQWs demonstrate similar absorption coefficients as the SQWs, but they can cover a broader working wavelength range by tuning the well and barrier thicknesses. Because the DQWs could allow decoupling of the ground and excited states in the different QWs, they afford larger gain coefficients compared to the SQWs and potentially are suitable for IR photon sources. The widely tunable transition energies and large absorption coefficients of the lead halide perovskites-based quantum well structures make them promising alternative candidates for IR applications.

ASSOCIATED CONTENT

**Supporting Information**.

The following files are available free of charge:

S1 Intersubband transition energy

S2 Absorption coefficients for SQW and DQW

S3 Scattering lifetime and optical gain.

AUTHOR INFORMATION

**Corresponding Author**

Xuedan Ma: xuedan.ma@anl.govACKNOWLEDGMENT

We acknowledge support from the National Science Foundation DMR Program under the award no. DMR-1905990. Use of the Center for Nanoscale Materials, an Office of Science user facility, was supported by the U.S. Department of Energy, Office of Science, Office of Basic Energy Sciences, under Contract No. DE-AC02-06CH11357. We thank S. Birner for assistance in the